\documentclass[fleqn,10pt]{wlscirep}
\usepackage[utf8]{inputenc}
\usepackage[T1]{fontenc}
\usepackage{xcolor}

\title{Universal image segmentation for optical identification of 2D materials}

\author[1]{Randy M. Sterbentz} 
\author[1]{Kristine L. Haley}
\author[1,*]{Joshua O. Island}
\affil[1]{Department of Physics and Astronomy, University of Nevada Las Vegas, Las Vegas, Nevada 89154, USA}
\affil[*]{jisland@physics.unlv.edu}

\begin{abstract}
Machine learning methods are changing the way data is analyzed. One of the most powerful and widespread applications of these techniques is in image segmentation wherein disparate objects of a digital image are partitioned and classified. Here we present an image segmentation program incorporating a series of unsupervised clustering algorithms for the automatic thickness identification of two-dimensional materials from digital optical microscopy images. The program identifies mono- and few-layer flakes of a variety of materials on both opaque and transparent substrates with a pixel accuracy of roughly 95\%. Contrasting with previous attempts, application generality is achieved through preservation and analysis of all three digital color channels and Gaussian mixture model fits to arbitrarily shaped data clusters. Our results provide a facile implementation of data clustering for the universal, automatic identification of two-dimensional materials exfoliated onto any substrate. 
\end{abstract}

\begin{document}

\flushbottom
\maketitle

\thispagestyle{empty}

\section{Introduction}
Image segmentation has become an essential technique in fields from medical imaging\cite{clarke-MRI1995, ji-Fuzzy2012, forouzanfar-Parameter2010} and autonomous driving\cite{feng-Deep2020} to robotic perception\cite{schwarz-RGBD2018} and image compression\cite{ratakonda-Lossless2002, sanderson-Image1995, liangshen-segmentationbased1997}. Through unsupervised segmentation of large data sets, trained algorithms can recognize and predict elements of new images. An appealing application of image segmentation is in the thickness identification of two-dimensional (2D) materials from their digital optical microscopy images. Current flake detection methods rely heavily on identification by trained researchers, a human-learning process, in which flakes are identified by their contrast difference on a substrate after significant trial and error. Automatic thickness identification would relieve this tedious, time-consuming screening process and possibly improve identification accuracy. 

The easiest implementation of image segmentation for 2D materials is by thresholding. This is performed by analyzing image contrast, from reflectance or transmittance for example, and partitioning regions of an image based on contrast level difference. This technique has been widely and successfully employed in the identification and characterization of exfoliated 2D materials\cite{li-Rapid2013, bing-Optical2018, taghavi-Thickness2019, ni-Graphene2007, jung-Simple2007, wang-Optical2009, zhang-Optical2017, yu-Investigation2017, wang-Thickness2012, rubio-bollinger-Enhanced2015, jessen-Quantitative2018}. Thresholding techniques, while easily implemented, suffer from inaccuracy when contrast differences become relatively small, for example a single layer of graphene on a silicon/silicon dioxide (Si/SiO$_2$) substrate, and can be highly dependent on precise experimental conditions hindering universal application.  

Recently, a variety of more advanced machine learning techniques have emerged to automate and improve the process of identifying exfoliated 2D materials\cite{yang-Automated2020, wu-twodimensional2019, masubuchi-Deeplearningbased2020, masubuchi-Classifying2019, lin-Intelligent2018, li-Rapid2019, greplova-Fully2020, cellini-Layer2019, hu2020rapid, dong20213d, aleithan2020toward}. These include techniques based on neural networks and data clustering but have been primarily applied to opaque substrates and almost entirely to standard Si/SiO$_2$. Transparent substrates are commonly used for exfoliation or experiments on 2D materials\cite{castellanos-gomez-Deterministic2014, island-Precise2016, lippert-Influence2017, nguyen-Visibility2020, chu-Linear2020, taghavi-Thickness2019}. A method that can be applied to identify the thickness of any material on any substrate is highly desirable. 

Here we present an open source program\cite{github} written in Python 3 to automatically identify the thickness of exfoliated 2D flakes which can be universally applied to different materials and substrates. We combine three well-established clustering techniques to form a training script to segment layers of a flake, manually label the layers, and then use that training to test thicknesses of other flakes. The program presents roughly a 95\% pixel accuracy for graphene and transition metal dichalcogenides on silicon/silicon dioxide and polydimethylsiloxane (PDMS) substrates. Importantly, no change to the program's adjustable parameters are needed to identify different materials on different substrates, allowing simple and universal application to any material/substrate combination.

\section{Overview of the program} 
\begin{figure}
\includegraphics[width=6.6 in]{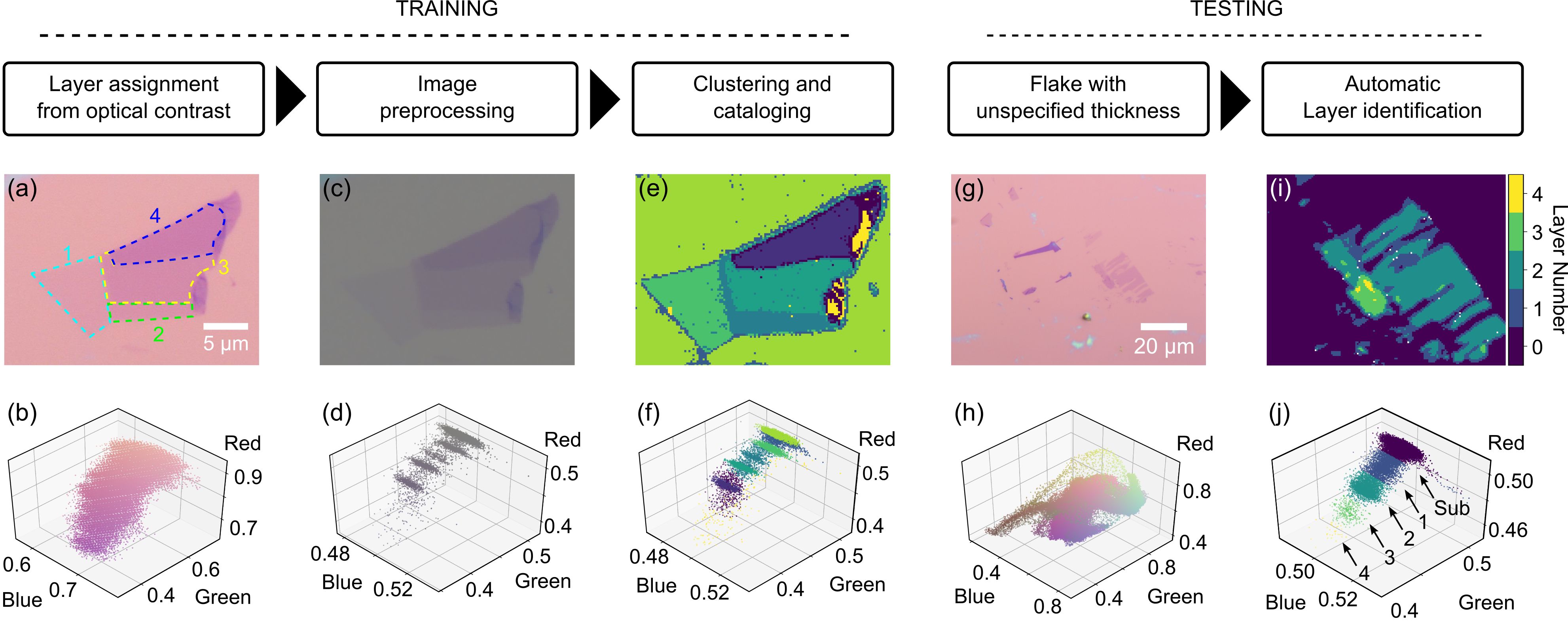}
\caption{\label{fig1} \textbf{Overview of the program composed of training and testing stages.}
(a-b) Raw optical image of a few-layer graphene flake on 300 nm of SiO$_2$ (a) and its corresponding scatter plot of each pixel in RGB color space (b). The scatter plot data points are colored to match their RGB value.
(c-d) The result after preprocessing the image in panel (a). The preprocessing reveals well-defined clusters in the associated scatter plot (d).
(e-f) A color plot of pixel-cluster association (e) and corresponding scatter plot (f). Each pixel is colored based on its most probable data cluster identity. 
(g-h) Raw optical image of few-layer graphene flake with unknown thickness (g) used for testing and its corresponding RGB scatter plot (h).
(i-j) Crop of panel (g) around the flake of interest (i) and corresponding scatter plot after the testing stage (j).
}
\end{figure}

 An overview of the program applied to graphene on Si/SiO$_2$ is shown in Figure \ref{fig1}. The training stage begins with a set of optical microscopy images cropped to few-layer flakes whose thicknesses are determined using optical contrast methods\cite{li-Rapid2013}. Figure \ref{fig1}(a) shows a cropped optical image of a few-layer graphene flake on a Si/SiO$_2$ (300 nm SiO$_2$) substrate with each layer thickness labeled (1-4 layers). A scatter plot of the red, green, and blue (RGB) channel values (normalized to a range 0-1) for each pixel in the image is shown in Figure \ref{fig1}(b). The data points have been colored according to their RGB values. At this stage, the scatter plot shows only broad features that can be generally associated with the substrate (pink) and few-layer graphene (purple) but no clear correspondence to individual thicknesses can be made. The raw image is preprocessed using a bilateral filter to reduce noise and a background normalization using a planar fit. The result, after compression to roughly 10,000 total pixels, is shown in Figure \ref{fig1}(c). Preprocessing reveals the individual clusters of data in the scatter plot (Figure \ref{fig1}(d)) associated with the substrate and flake layers. 

The location and distribution of these clusters in RGB space are found using a series of unsupervised clustering techniques summarized here and detailed further below. The centers are first located using mean shift and density-based spatial clustering. Once the centers are identified, fit characteristics such as the weight, mean position, and distribution of each cluster are found using a Gaussian mixture model. An image of the result of the fitting algorithm is shown in Figure \ref{fig1}(e). The pixels in the color plot have been colored according to the fit results and the scatter plot (Figure \ref{fig1}(f)) shows these fit assignments in more detail. Once the cluster characteristics are extrapolated for several training images, a master catalogue is created that ties the fit clusters to the predetermined flake thicknesses in our training images.

To determined accuracy of the training, we test the master catalogue on a set of images with identified thicknesses. An example of the testing results for graphene on Si/SiO$_2$ is shown in Figures \ref{fig1}(g-j). In the testing stage, untrained optical images (Figure \ref{fig1}(g-h)) are first preprocessed using the same procedure as in training but without cropping. The preprocessed images are then checked against the master catalogue for flake thickness assignment given the pixel location in RGB space. Figure \ref{fig1}(i) shows the result (cropped to show detail of flake of interest) with layer thicknesses identified by the color bar. The associated scatter plot in Figure \ref{fig1}(j) shows the corresponding clusters and layer assignments. In the following section we detail the implemenation of the clustering algorithms before presenting results of our program performed on other materials and substrates. 

% (a) Raw optical image of a few-layer graphene flake on 300 nm of SiO$_2$.
% (b) Scatter plot of the RGB color values for each pixel in (a). Pixel color matches their RGB value.
% (c) Same image after preprocessing, normalizing the substrate to gray.
% (d) Pixels of the image in (c) plotted in RGB space. The preprocessing reveals well-defined clusters in the scatter plot.
% (e) Trained image after clustering. The color assignment is arbitrary but indicates different layer thicknesses.
% (f) Scatter plot of the pixels in (e) colored by their layer assignment.
% (g) Raw optical image of a few-layer graphene flake used for testing.
% (h) Scatter plot of all the pixels in the image in panel (g).
% (i) Image of the same flake with the layers identified using the trained database.
% (j) A scatter plot showing the identified clusters for the test image (i) in RGB space.

\section{Unsupervised clustering algorithms} 

Our training script incorporates three clustering algorithms to identify the center of the data clusters and fit their distributions. Without explicitly knowing the number of clusters (layers) in the image, the script begins with an unsupervised method of determining the seed number. We use mean shift\cite{comaniciu-Mean2002, zhou-RegionBased2008, zhou-Mean2011} and density-based\cite{ester-densitybased1996, ruixu-Survey2005} algorithms to first find these cluster centers which are then fed to a Gaussian mixture model for fitting arbitrary ellipsoidal distributions.   

\begin{figure}
\centering
\includegraphics[width=3.3 in]{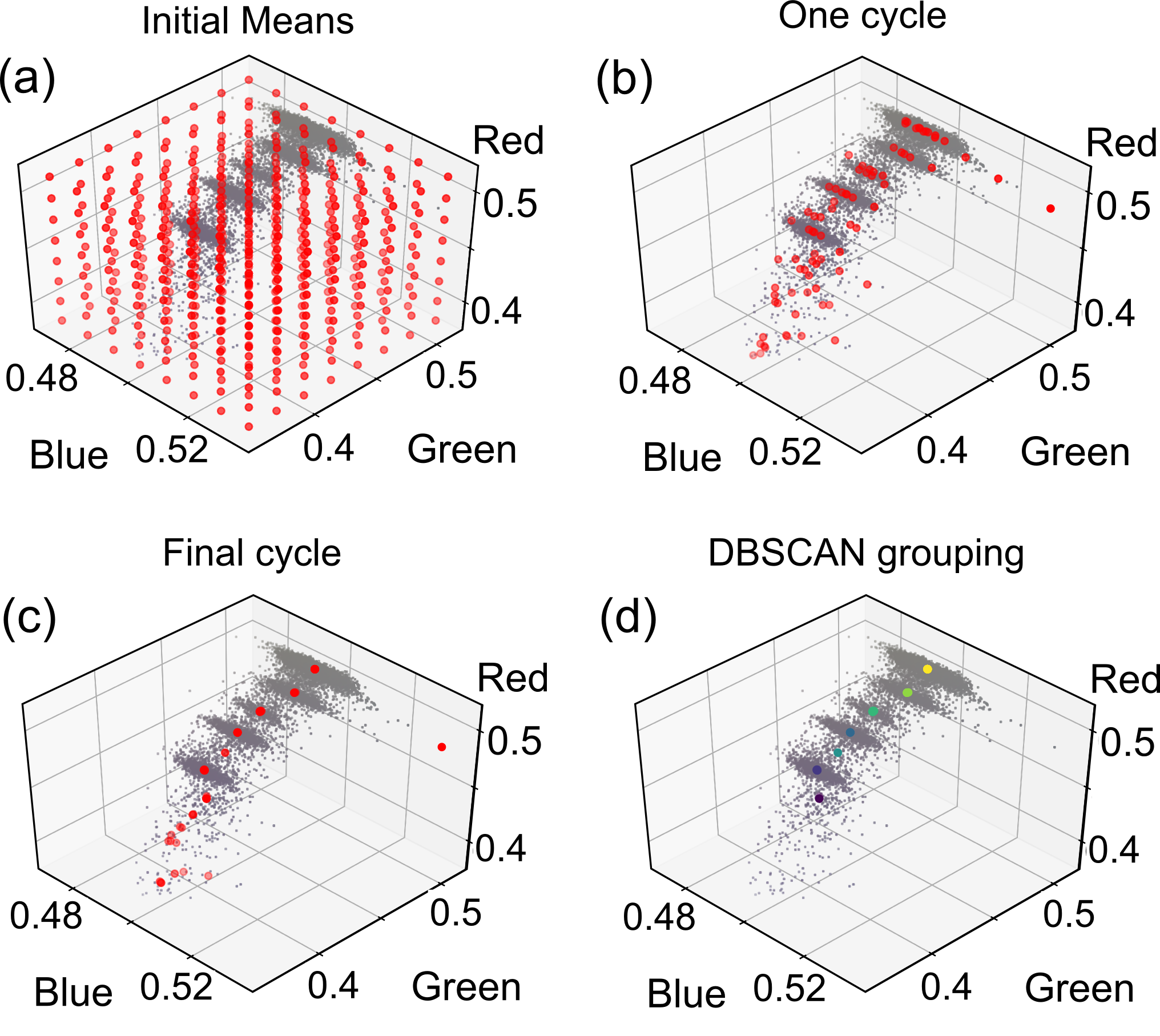}
\caption{\label{fig2} \textbf{Mean shift and DBSCAN clustering for identification of cluster centers.} 
(a) The same data as in Figure \ref{fig1}(d) with red closed circles showing the initial positions of the mean points in the mean shift algorithm.   
(b) Scatter plot after one cycle of mean shift; outlier mean points have been deleted and others have moved towards their local density maxima.  
(c) The final state of the mean points after they have converged to their maxima. 
(d) The RGB pixels plotted with the identified cluster centers (colored closed circles) after the DBSCAN algorithm.
}
\end{figure}

Mean shift is an unsupervised machine learning algorithm that locates centers of high data density. The algorithm begins by populating color space with an array of points, referred to as “mean points” ($\vec{\rho}_k$). Figure \ref{fig2}(a) shows the same data as in Figure \ref{fig1}(d) but with red closed circles indicating the initial positions of the equally spaced mean points (an $8\times8\times8$ array). The next step groups all data points within a defined radius ($\epsilon$) of each mean point together. We define $\epsilon$ to be just large enough to overlap with its nearest neighbors. The average location of the data pixels within $\epsilon$ of a given mean point becomes the new position of that mean point ($\vec{\rho}_k'$) after one iteration of the algorithm. This is calculated by:
\begin{equation}
    \vec{\rho}_k'=\frac{1}{M}\sum_i{\vec{x}_i} \textrm{ for } |\vec{x}_i-\vec{\rho}_k|<\epsilon, 
\end{equation} 
where $\vec{x}_i$ is the position of each data point and $M$ is the total number of data points within $\epsilon$ of $\vec{\rho}_k$. In this way, each mean point gradually shifts towards higher densities of data. Figure \ref{fig2}(b) shows one iteration of the algorithm. Several points have moved to their new mean positions according to the data within $\epsilon$ of each $\vec{\rho}_k$. 

Mean shift is computationally slow, having to calculate the distance between every data point ($\approx$10,000 pixels) and every mean point (initially 512). To increase efficiency, mean points that have no data within $\epsilon$ after the first cycle, and thus make no contribution towards a data cluster, are deleted. Additionally, mean points may approach their local maximum at different rates. Per mean point, as soon as the number of data points within their radius starts to decrease, they are turned off and no longer involved in future calculations. Figure \ref{fig2}(c) shows the final state of the algorithm where all mean points have converged to their local density maxima. After this, outliers are removed before moving to the next algorithm. 

Once mean shift is complete, several mean points will themselves be clustered in color space and some mean points will have converged to outliers. Due to the ellipsoidal shape of the clusters in RGB space after preprocessing, the mean points will tend to lie along lines. An efficient algorithm for grouping these lines is Density-Based Spatial Clustering of Applications with Noise (DBSCAN)\cite{ester-densitybased1996, ruixu-Survey2005}. DBSCAN groups data together by following the trajectory of nearby points. The algorithm starts by “visiting” a random mean point. A radius (we find $\epsilon/2$ works well) around it is checked for other mean points. If none are found, the starting point is labeled an outlier. If there are neighbors, they are grouped together. One of the other points in this group is visited next, checking the same radius around itself to find new points to add to the group. This repeats until no new points are added to the group and every point within the group has been visited. Once the group is finished, a new group starts at a randomly chosen mean point and the process repeats. The centers of each group are found by averaging their respective mean points. Figure \ref{fig2}(d) shows the result after running the DBSCAN algorithm on the mean points in Figure \ref{fig2}(c).

The combination of mean shift and DBSCAN presents an unsupervised method of determining how many clusters are in a given image and their centers. Following this step, this information can be used to seed a more powerful clustering technique for data with ellipsoidal distributions. The popular K-means clustering technique\cite{celebi-comparative2013, dhanachandra-Image2015, dhanachandra-Survey2017} for example is undesirable here as it assumes spherical clusters. Instead, we use a multivariate Gaussian Mixture Model (GMM)\cite{gupta-gaussianmixturebased1998, nguyen-Dirichlet2011, permuter-study2006, ribeiro-Hand2006, santosh-Tracking2013, ji-Fuzzy2012} that allows fitting of data with arbitrary normal distributions. This expands application of the program by automatically handling new materials and substrate combinations that may have different cluster distributions in RGB space.

In the GMM, each fitting ellipsoid has three characteristics developing throughout the process: the weight ($\phi_k$, defining the number of data points near ellipsoid $k$), the centroid ($\vec{\mu}_k$, defining the mean of the data points belonging to ellipsoid $k$), and the covariance matrix ($\Sigma_k$, defining the shape and orientation of ellipsoid $k$ in RGB space). These characteristics are used to calculate the probability $\gamma_{ik}$ of a data point ${\vec{x}_i}$ belonging to ellipsoid $k$. This probability is given by:
\begin{equation}
    \gamma_{ik}=\frac{\phi_k\mathcal{N}(\vec{x}_i,\vec{\mu}_k,\Sigma_k)}{\sum_{j=1}^K\phi_j\mathcal{N}(\vec{x}_i,\vec{\mu}_j,\Sigma_j)},
\end{equation}
where $K$ is the total number of clusters and $\mathcal{N}$ is the three-variable (for 3-dimensional RGB space) Gaussian distribution given by:
\begin{equation}
    \mathcal{N}(\vec{x}_i,\vec{\mu}_k,\Sigma_k)=[(2\pi)^3 |\Sigma_k| e^{(\vec{x}_i-\vec{\mu}_k)^T\Sigma_k^{-1}(\vec{x}_i-\vec{\mu}_k)}]^{-\frac{1}{2}}
\end{equation}
The weights, means, and covariance matrices used in these relations are calculated through:
\begin{equation}
    \phi_k=\frac{1}{N}\sum_{i=1}^N\gamma_{ik},
\end{equation}

\begin{equation}
    \vec{\mu}_k=\frac{\Sigma^N_{i=1}\gamma_{ik}\vec{x}_i}{\Sigma^N_{i=1}\gamma_{ik}},
\end{equation}

\begin{equation}\label{co}
    \Sigma_k=\frac{\sum_{i=1}^N\gamma_{ik}|\vec{x}_i-\vec{\mu}_k|^2}{\sum_{i=1}^N\gamma_{ik}}.
\end{equation}

\begin{figure}[ht!]
\includegraphics[width=3.37 in]{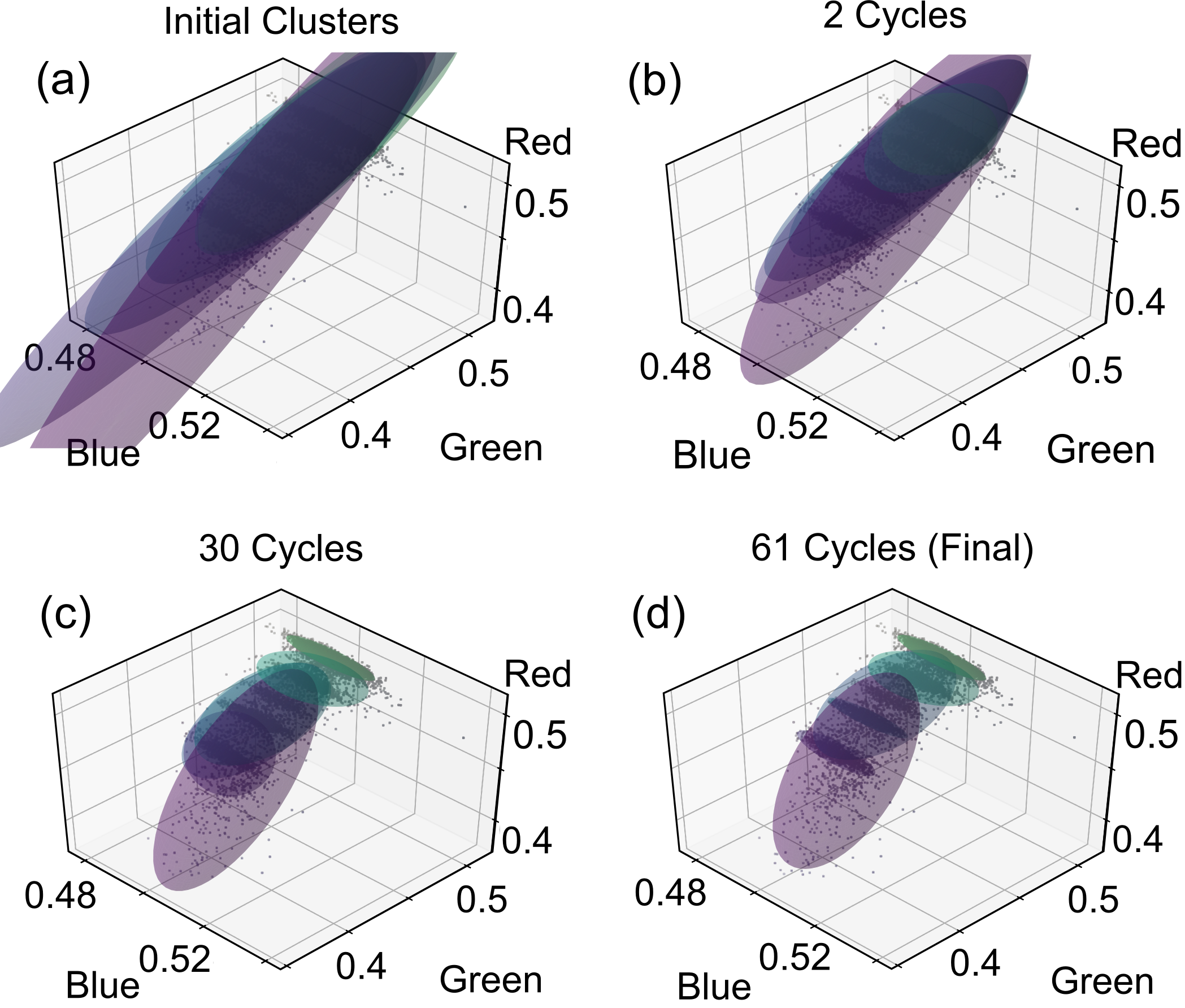}
\caption{\label{fig3} \textbf{Cluster fitting with a Gaussian mixture model (GMM).}
The scatter plot from Figure \ref{fig1}(d) is superimposed with 95\% confidence ellipsoids based on the fit characteristics of the GMM-EM algorithm. (a) The initialized ellipsoids show little correspondence to the underlying data. (b) After two cycles of expectation-maximization, the ellipsoids better resemble the data clusters. (c) After 30 cycles, some ellipsoids have nearly converged on their data clusters. (d) The convergence condition is reached after 61 cycles. 
}
\end{figure}

First we initialize each of the fitting ellipsoids by setting all initial weights to $1/K$. The centroids are taken directly from the results of DBSCAN $\vec{\mu}=\vec{\rho}$. The covariance matrices are initialized from the centroids using Equation \ref{co} with $\gamma_{ik}=1$. Figure \ref{fig3}(a) shows the initialization of the fitting ellipsoids for our example few-layer graphene data set from Figure \ref{fig1}(d) and Figure \ref{fig2}. The ellipsoids have been scaled to a 95\% confidence level.  

An unsupervised machine learning algorithm, referred to as expectation-maximization (EM), is used to further optimize the ellipsoid parameters and fit the data. The expectation step determines $\gamma_{ik}$ based on the initialized weights, centroids, and covariance matrices calculated above. The maximization step uses these probabilities to re-calculate each cluster’s weight, centroid, and covariance matrix. These two steps iterate and gradually the ellipsoid parameters converge. Figure \ref{fig3}(b) shows the algorithm results after 2 cycles and Figure \ref{fig3}(c) shows the results after 30 cycles. After 30 cycles, the ellipsoids resemble the distributions of the data with several small tight ellipsoids corresponding to the substrate and 1-4 layers of graphene, and two larger ellipsoids (purple and blue) accounting for noise. The max change of all cluster's weights between maximization steps ($\Delta\phi_k < 0.0001$) is used to define convergence and end the algorithm. Figure \ref{fig3}(d) shows the results of the algorithm after convergence (total 61 cycles) for this data set. Note that the large purple and blue ellipsoids are a product of over-fitting the data (fitting 7 ellipsoids to 5 data clusters). These ellipsoids do not contribute to the master catalogue, but are important for fitting data points associated with thicker layers ($>4$) and outliers. The over fitting also allows the primary ellipsoids to confine themselves to the core of their data clusters. Once convergence has been reached, only ellipsoids that fit well to known layer thicknesses are added to a catalogue.

The training process is repeated for multiple flakes of the same material and substrate (we trained around 5 for each material/substrate combination), saving their ellipsoid characteristics into the same catalogue. A master catalogue is then created by averaging together the characteristics of ellipsoids with like-thickness. This master catalogue is the tool with which we can test other images to determine their flake layer thicknesses (Figure \ref{fig1}(i-j)). 

% Next several comments are pre-Randy.
%
% (a) Scatter plot from Figure \ref{fig1} (d) with cluster ellipoids super-posed on top. The  ellipsoids indicate the location and 95\% confidence interval of the initial fitted clusters in the GMM algorithm (coloring is arbitrary). 
% (b) The RGB pixels (grey) and the fitted clusters from the GMM algorithm after 36 cycles. 
% (c) The RGB pixels (grey) and the fitted clusters from the GMM algorithm after 70 cycles.
% (d) The RGB pixels (grey) and the fitted clusters from the GMM algorithm after convergence and 106 total cycles. Note that one cluster has been removed for clarity.

\section{General application to other materials and substrates}

Our script can be universally applied to the identification of other 2D material thicknesses on opaque and transparent substrates. This generality is achieved by analyzing all three dimensions of the color-space data and fitting the resulting clusters of arbitrary shape with our GMM-EM algorithm. Importantly, no change in the adjustable parameters ($\epsilon$ or GMM convergence) are required for the following results.

\begin{figure}
\includegraphics[width=6.6 in]{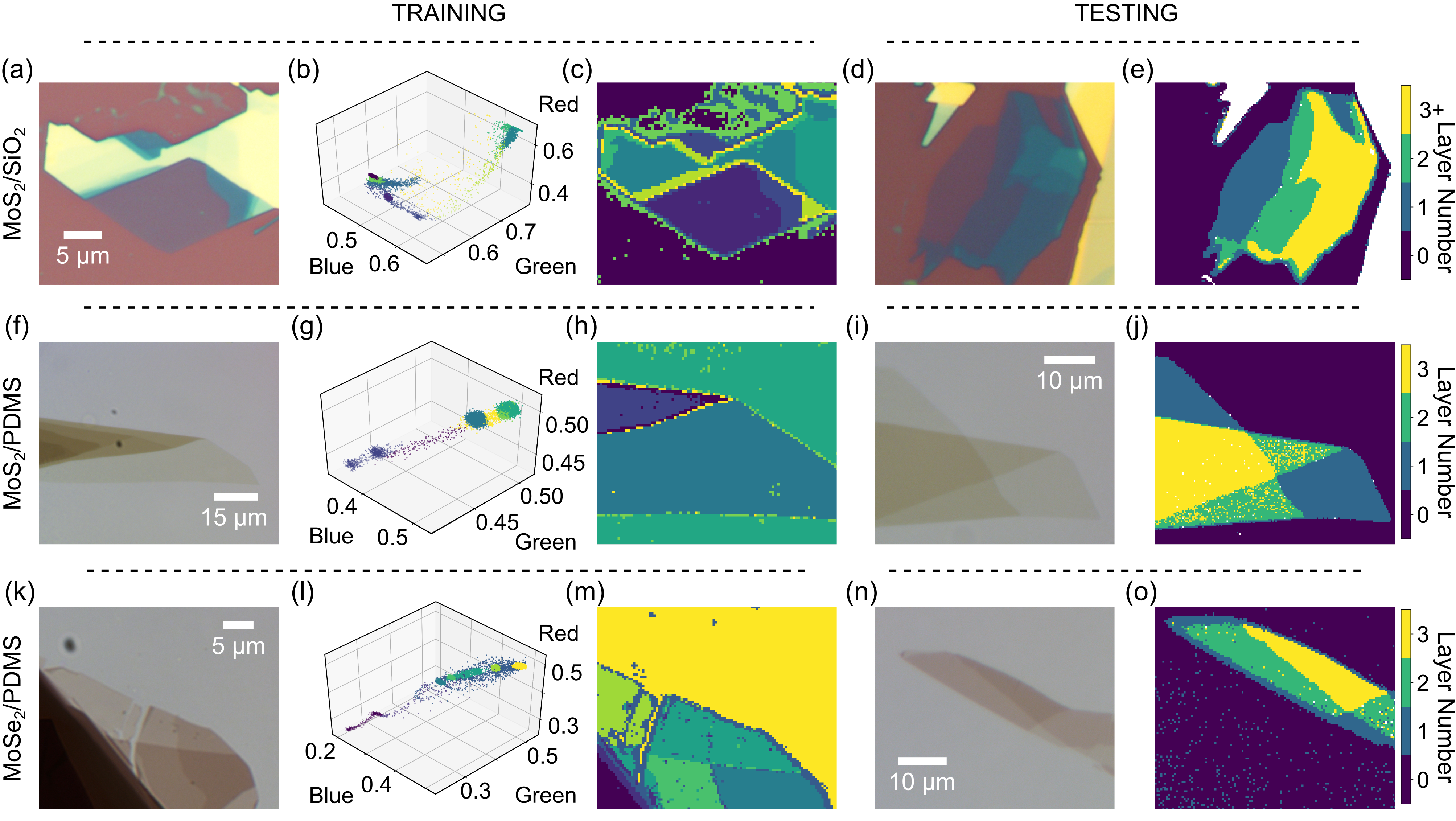}
\caption{\label{fig4} \textbf{General thickness identification of 2D materials on opaque and transparent substrates.}
(a-c) An example training process for MoS$_2$ flakes exfoliated onto Si/SiO$_2$ substrates. (a) Raw optical image before preprocessing. (b) RGB scatter plot of the identified clusters with pixels colored according to the cluster they belong to. (c) Reconstructed image of the MoS$_2$ flake in (a) after the training step. 
(d-e) Testing process for MoS$_2$ on Si/SiO$_2$. (d) Raw optical image of an MoS$_2$ flake on Si/SiO$_2$. (e) Layer identification after testing the image in (d).    
(f-h) An example training process for MoS$_2$/PDMS.  
(i-j) An example testing process for MoS$_2$/PDMS. 
(k-m) An example training process for MoSe$_2$/PDMS. 
(n-o) An example testing process for MoSe$_2$/PDMS.  
}
\end{figure}

Figure \ref{fig4} displays the power of this generality by identifying the layer thickness of two additional materials, molybdenum disulfide (MoS$_2$) and molybdenum diselenide (MoSe$_2$), on opaque (Si/SiO$_2$) and transparent (polydimethylsiloxane (PDMS)) substrates. MoS$_2$ on Si/SiO$_2$ (Figure \ref{fig4}(a-e)) presents clusters very similar to those of graphene on Si/SiO$_2$ but they are separated further in RGB space (Figure \ref{fig4}(b)). Further training for this material/substrate combination would improve our testing results which only identify layer thicknesses of 1 and 2 (Figure \ref{fig4}(d-e)). From the covariance matrices we note that while all the data clusters are technically triaxial ellipsoids (none of the semi-axes are equal), the clusters for materials on Si/SiO$_2$ are roughly prolate spheroids with one axis (blue) an order of magnitude larger than the other two semi-axes (red and green). 

MoS$_2$ on PDMS (Figure \ref{fig4}(f-j)) presents clusters again extending along the blue axis, though not as strongly as materials on Si/SiO$_2$. The clusters are similarly well-separated in RGB space as they are for MoS$_2$ on Si/SiO$_2$. Testing for this set identifies monolayer, bilayer and trilayer thicknesses (Figure \ref{fig4}(j)). Finally, MoSe$_2$ on PDMS presents the most spherical ellipsoids of our investigation still slightly extending along blue, (Figure \ref{fig4}(k-o)) and mono- through trilayer thicknesses are easily identified (Figure \ref{fig4}(o)).  

\section{Discussion} 
Our investigation focuses on the development of a program that can be universally applied to different 2D materials and substrates. This requirement invariably introduces computation time when compared with other recent segmentation methods\cite{masubuchi-Deeplearningbased2020, han-DeepLearningEnabled2020}. The training time, for example, reported in Ref. \cite{han-DeepLearningEnabled2020} for the entire program is roughly 31 hours. Computation times for the training stage of our program depend on the image composition. A single layer image can take about ten minutes but images with multilayer flakes (more clusters) take as long as 5 hours. Our program results here are from training sets of roughly 10 images corresponding to about 10 hour computation time. However, this is a single event time cost because once the master catalogue is trained for a particular material and substrate combination, it can then be used repeatedly in the testing step, which is more efficient. 

Image testing requires roughly one minute to identify layer thicknesses of new images. Computation time is sufficiently short for testing because image pixels are simply compared with the master catalogue. This time would allow in-situ identification of flakes from images taken by human inspection of a substrate. The time spent scanning between images can take several minutes. The time may also be sufficient for an automated scanning system such as that presented in ref. \cite{masubuchi-Autonomous2018}. Improvements in computation time may be sought through further image compression or possibly reducing the testing step to two dimensions of the three-dimensional RGB space, possibly blue and either green or red, similar to algorithms presented in ref. \cite{masubuchi-Classifying2019}. Although, the dropped color dimension would have to be identified for a particular material/substrate combination. 

For each material/substrate combination investigated in this study, the pixel accuracy was determined by creating a ground truth image and comparing it, pixel-by-pixel, with the testing images (see Figure S1 in the supplemental materials for details). Pixel accuracy was slightly better for materials on PDMS but overall the program achieves an average accuracy of 95\% for the materials and substrates investigated in this study. This pixel accuracy is comparable to that achieved in studies based on much larger training sets. Ref. \cite{han-DeepLearningEnabled2020} reports pixel accuracy of 97\% from a training set of 917 images. Based on these results, normalized confusion matrices for each combination were calculated showing the individual layer accuracy as well. Finally, we note that a clear advantage to our approach is the simplicity of our program which relies on well-known and proven clustering techniques with relatively high pixel accuracy from small training sets.

\section{Conclusion} 

Summarizing, we have presented a code for the automatic identification of flake thicknesses that can be universally applied to a variety of 2D materials and substrates. The algorithm analyzes data clusters in RGB space of preprocessed optical microscopy images. It can accurately identify mono- and few-layer thicknesses with a pixel accuracy of 95\%. We anticipate the program will be of use for a wide variety of materials and substrates for the continued interest and investigation into the properties and characteristics of 2D materials. 

\section{Acknowledgements} 
The authors thank Jeffery Cloninger for technical support and Najme S. Taghavi and Dr. Andres Castellanos-Gomez for optical images of materials on PDMS.

\section{Author contributions statement}

J.O.I. conceived the project. R.M.S. coded the program with guidance from J.O.I.. R.M.S. and K.L.H. provided optical images of 2D materials. All authors wrote and reviewed the manuscript. 

%\twocolumngrid

\clearpage
\section{References}
\bibliography{Bib}

\clearpage
\section{Supplemental Materials: Universal image segmentation for optical identification of 2D materials}
\setcounter{equation}{0}
\setcounter{figure}{0}
\setcounter{table}{0}
\setcounter{page}{1}
\setcounter{section}{0}
\renewcommand{\thefigure}{S\arabic{figure}}
\renewcommand{\thesubsection}{S\arabic{subsection}}
\renewcommand{\theequation}{S\arabic{equation}}
\renewcommand{\thetable}{S\arabic{table}}
\renewcommand{\thepage}{S-\arabic{page}}
\section{Python code and packages} 
The software was written in Python 3.7.3 and includes the following packages: scipy==1.5.2, matplotlib==3.0.3, numpy==1.16.2, and opencv-python==4.2.0.34. The full Python code is available at https://github.com/islandlab-unlv/Universal-image-segementation.

\section{Pixel accuracy and confusion matrices}
\begin{figure*}[h]
\includegraphics[width=\linewidth]{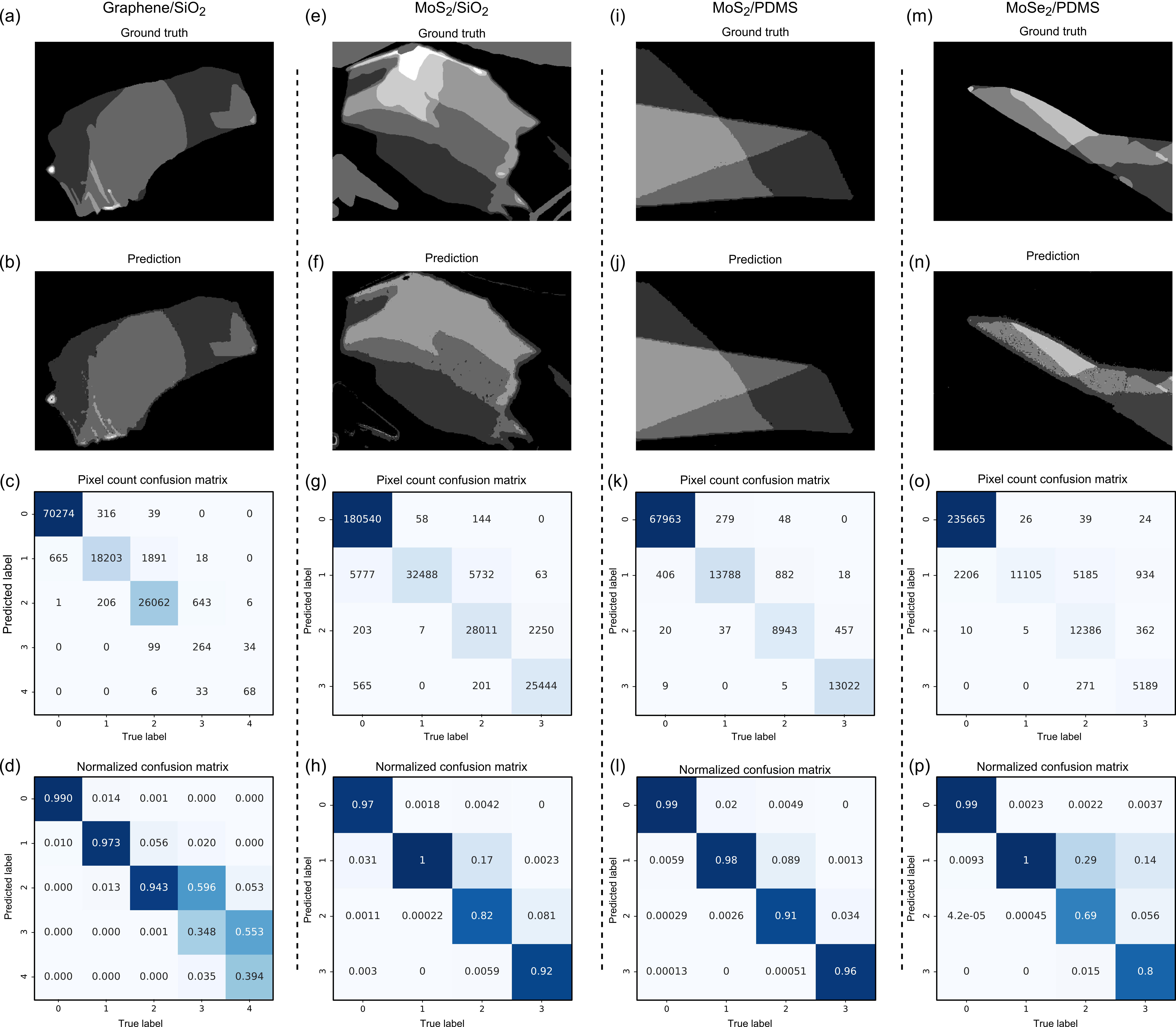}
\caption{
\textbf{Pixel accuracy testing.} 
(a-d) Ground truth (a) and prediction (b) images from testing for a graphene flake on a Si/SiO$_2$ substrate. Raw pixel count confusion matrix for graphene on SiO$_2$ (c), and the same confusion matrix normalized by column (normalized to true label) (d).
(e-h) Ground truth (e) and prediction (f) images from testing for a MoS$_2$ flake on a Si/SiO$_2$ substrate. Raw pixel count confusion matrix for MoS$_2$ on SiO$_2$ (g), and the same confusion matrix normalized by column (normalized to true label) (h).
(i-l) Ground truth (i) and prediction (j) images from testing for a MoS$_2$ flake on a PDMS substrate. Raw pixel count confusion matrix for MoS$_2$ on PDMS (k), and the same confusion matrix normalized by column (normalized to true label) (l).
(m-p) Ground truth (m) and prediction (n) images from testing for a MoSe$_2$ flake on a PDMS substrate. Raw pixel count confusion matrix for MoSe$_2$ on PDMS (o), and the same confusion matrix normalized by column (normalized to true label) (p).
}
\label{figS1}
\end{figure*}

In order to measure the fidelity of our program, we created confusion matrices by comparing the testing results (predictions) with manually recolored images based on layer thickness (ground truths). Figure \ref{figS1}(a) shows a ground truth image for a graphene flake on Si/SiO$_2$ and Figure \ref{figS1}(b) shows the testing result using the same colorscale. Figure \ref{figS1}(c) shows the raw counts for the confusion matrix of graphene on Si/SiO$_2$. Element $ij$ in the confusion matrix is the number of pixels predicted to be $i$-layer that are known to be $j$-layer. The diagonal thus is where the program predicted the layer thickness correctly. We calculate the pixel accuracy by taking the ratio of the trace of the matrix with the sum of all elements in the matrix. For graphene on SiO$_2$, we calculate a pixel accuracy of 96.7\%. For the other test samples in Figure \ref{figS1}, we calculate the following pixel accuracies: MoS$_2$/SiO$_2$ in Figure \ref{figS1}(e-h) (94.7\%), MoS$_2$/PDMS in Figure \ref{figS1}(i-l) (98.0\%), MoSe$_2$/PDMS in Figure \ref{figS1}(m-p) (96.7\%). 

Normalizing a confusion matrix by the true labels elucidates how well the program can predict each layer thickness. Columns 0-2 of Figure \ref{figS1}(d), for example, state we can successfully identify 99.0\% of the Si/SiO$_2$ substrate, 97.3\% of the monolayer graphene regions, and 94.3\% of the bilayer graphene regions. It appears to falter with tri- and quadlayer identification, but this is a result of the necessary preprocessing on the test image. Despite qualities of bilateral filtering as an edge-preserving filter, we always had some blurring at the flake-thickness boundaries while applying it. As a result, narrow strips of a layer thickness could be blurred away, as is the case with the thin stripes of trilayer present in the graphene/SiO$_2$ test image (light grey stripes at the bottom left of panels (a) and (b) of Figure \ref{figS1}). These confusion matrices were created off of a single test image per material/substrate tested, and are beholden to the biases that arise from those individual samples. With finer tuning of the bilateral filter and a larger test set, this blurring should be mitigated.

\end{document}